\documentstyle[epsf,eqsecnum,aps]{revtex}

\def\abstract#1{\vskip 7mm 
        \begin{center}{\large Abstract}\par \smallskip
                \begin{minipage}[c]{12cm}
                        \small #1
                \end{minipage}
        \end{center}
}
\def\title#1{\begin{center}{\Large\bf #1}\end{center}}
\def\author#1{\vskip 5mm \begin{center}{#1}\end{center}}
\def\address#1{\begin{center}{\it #1}\end{center}}

\def\be{\begin{equation}}
\def\ee{\end{equation}}
\def\bea{\begin{eqnarray}}
\def\eea{\end{eqnarray}}

\def\L#1{{\stackrel{(L)}{#1}}}
\def\*#1{{#1}^{*}{}}
\def\3{{}^3\!\!\,}
\def\g#1{\gamma_{#1}}

\def\K#1{K_{#1}}
\def\grr{\gamma_{rr}}
\def\gthth{\gamma_{\theta\theta}}
\def\Krr{K_{rr}}
\def\Kthth{K_{\theta\theta}}
\begin{document}
\begin{flushright}
TIT/HEP-376/COSMO-84 \\
\today
\end{flushright}
%%%%%%%%%%%%%%%%%%%%%%%%%%%%%%%%%%%%%%%
%                Title                %
%%%%%%%%%%%%%%%%%%%%%%%%%%%%%%%%%%%%%%%

\title{
Renormalization Group Approach to Einstein Equation 
in Cosmology
}
\author{
Osamu Iguchi\footnote{JSPS fellow,\quad E-mail:
  osamu@th.phys.titech.ac.jp}, 
Akio Hosoya\footnote{E-mail: ahosoya@th.phys.titech.ac.jp}
and Tatsuhiko Koike\footnote{E-mail: koike@rk.phys.keio.ac.jp}\\
}
\address{
        Department of Physics,  
        Tokyo Institute of Technology, \\ 
        Oh-Okayama Meguro-ku, Tokyo 152, Japan
}
\address{
        \ddag Department of Physics,
        Keio University, \\
        Hiyoshi, Kohoku, Yokohama 223, Japan
}
\abstract{
The renormalization group method has been adapted to the analysis 
of the long-time behavior of non-linear partial differential equation 
and has demonstrated its power in the study of critical phenomena 
of gravitational collapse. 
In the present work we apply the renormalization group 
to the Einstein equation in cosmology and carry out detailed 
analysis of renormalization group flow in the vicinity of 
the scale invariant fixed point in the spherically symmetric 
and inhomogeneous dust filled universe model.
}

\begin{minipage}[t]{5in}
\end{minipage}

%%%%%%%%%%%%%%%%%%%%%%%%%%%%%%%%%%%%%%%
%             Introduction            %
%%%%%%%%%%%%%%%%%%%%%%%%%%%%%%%%%%%%%%%
\section{Introduction}
\label{sec:intro}

Recently the renormalization group (RG) idea is applied to study 
the long-time asymptotics of the nonlinear partial differential 
equations \cite{BKL,CGO}.
The RG transformation there is 
integration of the equation up to a finite time 
followed by a rescaling of the dependent and independent variables.
The RG transformation together with the original differential
equation gives a RG equation.
Using the RG transformation, the problem on an infinite time is reduced 
to the problem on a finite time.
A fixed point of the RG transformation corresponds to a scale
invariant solution of the differential equation.
We can obtain the long-time behavior of the equation
by studying the flow around fixed points.

As an application of this RG method to the system of gravity, 
Koike, Hara and Adachi \cite{HKA} analyzed the Einstein 
equation to understand the problem of critical behavior of black hole 
mass in gravitational collapse found by numerical study \cite{Cho}.
A pedagogical exposition of the RG method in the deterministic 
system is given by Tasaki \cite{TH} in a simple but very illustrative 
example of motion of a point particle in the Newtonian gravity. 

Here we apply the RG to the Einstein equations in the cosmological situation.
For simplicity, we shall consider only two cases. 
One is a homogeneous and isotropic 
universe filled with a perfect fluid and the other is a spherically 
symmetric universe filled with dust. 
We shall study the flow near the fixed points of the RG equations, which 
have self-similarity.

The astronomical observations indicate that the present universe has
the hierarchical structure such as galaxies, clusters of galaxies, and 
super clusters, and that 
the two point correlation function of the galaxies and of 
the clusters of galaxies can be expressed roughly by a single 
power law \cite{COL}. 
The scale invariant Harrison-Zel'dovichi spectrum for the primordial density
perturbation has been successful in the study of structure formation of the
universe \cite{HZ}. 
These suggest that the present universe has some self-similarity 
and that the scale invariant solution plays an important role 
in cosmology.

In this paper we apply the renormalization group method 
to the Einstein equations in the cosmological context.
In Sec.II, we illustrate the application of the RG method 
to the heat equation with the nonlinear term. 
In Sec.III, we apply the RG method to the Einstein equations.
Section IV is devoted to the summary and discussions.

%%%%%%%%%%%%%%%%%%%%%%%%%%%%%%%%%%%%%%%
%           RG transformation         %
%%%%%%%%%%%%%%%%%%%%%%%%%%%%%%%%%%%%%%%
\section{Renormalization group transformation
----- heat equation with nonlinear term -----}
\label{sec:rgt}

In this section, we review the RG method for 
nonlinear partial differential equations \cite{BKL}.
First we consider the heat equation with the nonlinear term
as a simple example:
\bea
   \frac{\partial u(x,t)}{\partial t} &=& \frac{1}{2}\left[ 
        u''(x,t)+\lambda u^2(x,t)\right],
\label{eq:heat}
\eea
where the prime denotes the spatial derivative and $\lambda$ is a coupling 
constant.
The equation (\ref{eq:heat}) has scale invariance 
under the following scale transformation:
\bea
   x &\longrightarrow& Lx,
\nonumber \\
   t &\longrightarrow& L^2 t,
\nonumber \\
   u(x,t) &\longrightarrow& L^2 u(Lx,L^2t),
\label{eq:st}
\eea
where $L$ is a parameter of the scale transformation and is taken to 
be larger than 1. 
Namely, if $u(x,t)$ is a solution of Eq.(\ref{eq:heat}), 
the scaled function, 
\bea
   \L{u}(x,t) &=& L^2 u(Lx,L^2t),
\label{eq:st_u}
\eea
is also a solution of Eq.(\ref{eq:heat}).
We can thus obtain a one-parameter family of solutions, provided 
that $u(x,t)$ is a solution.

Here we define the RG transformation ${\cal R}_L$ of a function 
of $x$ by
\bea
   {\cal R}_L u(x,1) &=& \L{u}(x,1).
\label{eq:def_rg}
\eea
In short, the ${\cal R}_L$ is a map from a set of initial data to
another.
It is convenient to take the initial time to be $t=1$.
The RG transformation ${\cal R}_L$ has a semi-group property:
\bea
   {\cal R}_{L^n} &=& {\cal R}_{L^{n-1}} \circ {\cal R}_L.
\eea
Letting $t=1$ and then $L^2=t$ in Eqs.(\ref{eq:st_u}) and 
(\ref{eq:def_rg}), 
we can express an arbitrary solution $u(x,t)$ as an initial data
$u(\cdot,1)$ transformed by RG transformations:
\bea
   u(x,t) &=& t^{-1}\stackrel{(t^{1/2})}{u}(xt^{-1/2},1).
\label{eq:ru}
\eea
The large $L$ means the late time.
Repeating the RG transformation (\ref{eq:def_rg}), we can see the 
long-time behavior of the solution $u(x,t)$ in Eq.(\ref{eq:heat}).

Denoting $L=e^\tau$, 
we have from Eq.(\ref{eq:st_u}) that
\bea
   \frac{d\L{u}}{d\tau} = L\frac{d\L{u}}{dL}
   = 2\L{u}+x\L{u'}+2\frac{\partial\L{u}}{\partial t}.
\label{eq:prerg_heat}
\eea
Using the original partial differential equation (\ref{eq:heat}) we have
\bea
   \frac{d\L{u}}{d\tau} &=& 2\L{u}+\lambda\L{u^2}+x\L{u'}+\L{u''}.
\label{eq:rg_heat}
\eea
This is the equation satisfied by the scaled function $\L{u}$, which
we call the RG equation. 
We note that the equation (\ref{eq:rg_heat}) has no explicit scale $L$ 
dependence because of the scale invariance of the original equation 
(\ref{eq:heat}).

%%%%%fixed point%%%%%

We investigate the fixed point of the RG equation (\ref{eq:rg_heat}).
The fixed point $\*{u}$ is defined by 
\bea
  {\cal R}_L\*{u} &=& \*{u}
\label{def_fp} 
\eea
for any $L>1$.
This condition means that the field profile is unchanged after time
evolution followed by suitable rescaling.
In general this condition is equivalent to
\bea
   \frac{d\L{\* u}}{d\tau} &=& 0.
\label{eq:con_fp}
\eea
{}From Eq.(\ref{eq:rg_heat}), $\*{u}$ satisfies the following equation,
\bea
   2\*{u}+\lambda\*{u}^2+x\*{u}'+\*{u}'' &=& 0.
\label{eq:eq_fp}
\eea

In the homogeneous case, we can easily obtain the fixed points:
\bea
   \*{u} &=& 0 \quad \mbox{and} \quad -\frac{2}{\lambda}.
\label{eq:sol_fp}
\eea

%%%%%perturbation%%%%%

To investigate the character of the fixed point Eq.(\ref{eq:sol_fp}), 
we consider the linear perturbation around the fixed point 
Eq.(\ref{eq:sol_fp}).
The perturbed quantity $\L{\delta u}$ is defined by 
\bea
   \L{u} &=& \*{u}+\L{\delta u},
\label{eq:def_perh}
\eea
where $\L{\delta u}$ is assumed small.
Substituting Eq.(\ref{eq:def_perh}) into Eq.(\ref{eq:rg_heat}) and 
neglecting the second order term $\L{\delta u^2}$, 
we obtain the linearized equation for $\L{\delta u}$:
\bea
   \frac{d\L{\delta u}}{d\tau} &=& \left\{ 
   \begin{array}{cl}
      2\L{\delta u}+x\L{\delta u'}+\L{\delta u''} & (\*{u}=0) \\
      -2\L{\delta u}+x\L{\delta u'}+\L{\delta u''} & 
      (\*{u}=-\frac{2}{\lambda})  
   \end{array}
   \right. .
\label{eq:per_heat}    
\eea
We require the boundary condition
\bea
   \L{\delta u} \longrightarrow 0 && \quad (|x| \longrightarrow \infty) .
\eea
We are going to find the normal modes with the ansatz:
\bea
   \L{\delta u} &=& f(x)e^{-\frac{x^2}{2}+\omega\tau},
\label{eq:per_ansatz}
\eea
where $f(x)$ is a function to be determined below and $\omega$ is a constant.
{}From Eqs.(\ref{eq:per_heat}) and (\ref{eq:per_ansatz}), we have 
\bea
   f''-xf'-(\omega-1)f &=& 0 \quad (\*{u}=0),
\nonumber \\
   f''-xf'-(\omega+3)f &=& 0 \quad (\*{u}=-\frac{2}{\lambda}).
\label{eq:f_heat}
\eea
The regularity at $x=0$ and the boundary condition at $|x|=\infty$ imply 
\bea
   f(x) &=& H_n(x),
\\
   \omega &=& \left\{
       \begin{array}{cl}
           1-n & (\*{u}=0) \\
           -3-n & (\*{u}=-\frac{2}{\lambda})
       \end{array}
   \right. ,
\label{eq:sol_per_heat}
\eea
where $H_n(x)$ is the Hermite polynomial and $n=0,1,2,\cdots$.

{}From Eq.(\ref{eq:sol_per_heat}), $\*{u}=-2/\lambda$ is an attractor 
because all $\omega$'s are negative.
On the other hand, $\*{u}= 0$ has 
only one relevant mode ($n=0$).

%%%%%long-time behavior%%%%%

We can discuss the long-time behavior of a solution of the nonlinear 
diffusion equation (\ref{eq:heat}), if $u(x,1)$ is sufficiently close 
to the self-similar profile, $u^{*}$. 
{}From Eq.(\ref{eq:sol_fp}), we obtain the two self-similar profiles. 
Suppose the initial spatial profile of $\L{\delta u}$ is expressed 
as a superposition of the normal modes $H_{n}e^{-x^2/2}$. 
As we have seen from Eq.(\ref{eq:sol_per_heat}), 
if $u(x,1)$ is sufficiently close to the fixed point $u^{*}=-2/\lambda$, 
the solution approaches to $-2t^{-1}/\lambda$ in course of time 
because all modes of perturbation are irrelevant. 
On the other hand, there is the only one growing mode ( $n=0$ ) of 
perturbation around the fixed point $u^{*}=0$.
As time goes on, the behavior of the solution near this fixed point 
is dominated by the relevant mode $n=0$.
This relevant mode corresponds a gaussian distribution $t^{-1/2}e^{-x^2/(2t)}$.

{}From this instructive example, we see that if the perturbation around 
the fixed point has a finite number of relevant modes or no relevant modes, 
we have some prediction power for the long-time behavior of
nonlinear partial differential equation.

%%%%%%%%%%%%%%%%%%%%%%%%%%%%%%%%%%%%%%%
%         Renormalization group       %
%         for Einstein equation       %
%%%%%%%%%%%%%%%%%%%%%%%%%%%%%%%%%%%%%%%
\section{Renormalization group for Einstein equation}
\label{sec:rge}

In this section, we apply the RG method, which is explained 
in the previous section, to the Einstein equations.

Take a synchronous reference frame where the line element is
\bea
   ds^2 &=& g_{\mu\nu}dx^{\mu}dx^{\nu} \nonumber \\
        &=& -dt^2+\gamma_{ij}(x^k,t)dx^idx^j . 
\label{eq:metric}
\eea
Throughout this paper Latin letters will denote spatial indices and
Greek letters spacetime indices. 
The matter is taken to be a perfect fluid characterized by
the energy-momentum tensor 
\be
	T_{\mu\nu}=(\rho + p)u_{\mu}u_{\nu} + p g_{\mu\nu} , 
\label{eq:matter}
\ee
where $p$, $\rho$, and $u_{\mu}$ 
are pressure, energy density, and four-velocity, respectively. 
We assume that the equation of state of the fluid is 
\bea
 p &=& ({\Gamma} -1) \rho,
\eea
where $\Gamma$ is a constant. 
The Einstein equations are  
\bea
  \dot{\gamma}_{ij} &=& 2K_{ij},
\label{eq:K} \\
  \dot{K}_{ij} &=&
  	-\3 R_{ij}-KK_{ij}+2K_i^l K_{lj}
  	+\frac{\kappa\rho}{2}[2\Gamma u_i u_j+(2-\Gamma)\gamma_{ij}],   
\label{eq:ein1}\\
  \kappa\rho &=&  \frac{\3 R+K^2-K_l^m K_m^l}{2(1+\Gamma u_lu^l)}, 
\label{eq:ein2} \\
  \kappa\Gamma\rho u_i 
	 &=& -\frac{1}{\sqrt{1+u_lu^l}} (K_{i;j}^j-K_{, i}) , 
\label{eq:ein3}
\eea
where $K_{ij}$ is the extrinsic curvature, 
$\3 R_{ij}$ is the Ricci tensor associated with $\gamma_{ij}$,
and $\kappa \equiv 8\pi G$. 
A dot denotes the derivative with respect to $t$, a semicolon denotes
the covariant derivative with respect to $\gamma_{ij}$.

Hereafter we consider the RG transformation 
for the dynamical variables $\g{ij}$ and $K_{ij}$.
In the following subsections, we investigate the two cases; 
the one is a homogeneous and isotropic universe, 
the other is a spherically symmetric inhomogeneous dust universe.

%%%%%%%%%%%%%%%%%%%%%%%%%%%%%%%%%%%%%%%
%    Homogeneous and Isotropic case   %
%%%%%%%%%%%%%%%%%%%%%%%%%%%%%%%%%%%%%%%
\subsection{Homogeneous and Isotropic case}
\label{sec:homo}

%%%%% RG equation %%%%%

We consider the homogeneous and isotropic universe as a simple case.
This case is rather trivial because the field equation becomes an 
ordinary differential equation.
Nonetheless, this gives a nice warming up model to familiarize us 
to the RG approach to the universe.
In this case, the spatial metric is written by 
\bea
   \gamma_{ij}(x^k,t) &=& 
	 \frac{a^2(t)}{\left(1+\frac{k(t)r^2}{4}\right)^2}\delta_{ij}.
\label{eq:homometric}
\eea
where $r^2\equiv\delta_{ij}x^ix^j$, 
and $a(t)$ and $k(t)$ are the functions of time $t$ to be studied.
Substituting Eq.(\ref{eq:homometric}) into 
Eqs.(\ref{eq:K}) -- (\ref{eq:ein3}), we get 
\bea
   \dot{a} &=& aH, 
\label{eq:def-H} \\
   \dot{H} &=& -3H^2-\frac{2k}{a^2}+\frac{2-\Gamma}{2}\kappa\rho,
\label{eq:H} \\
   \dot{k} &=& 0,
\label{eq:homok} \\
   \kappa\rho &=& 3\left[H^2+\frac{k}{a^2}\right],
\label{eq:homo2} 
\eea
where $H$ is the Hubble parameter.
{}From the conservation law of the energy density, we have 
\bea
   \kappa\rho &=& M a^{-3\Gamma},
\label{eq:rho_homo}
\eea
where $M$ is an arbitrary constant.

First, we consider the following scale transformation.
\bea
   t   &\longrightarrow& Lt, \nonumber \\
   a(t) &\longrightarrow& \L{a}(t)\equiv L^{-2/(3\Gamma)}a(Lt), \\
   k(t) &\longrightarrow& \L{k}(t)\equiv L^{2(3\Gamma-2)/(3\Gamma)}k(Lt),
\label{eq:rgk_homo}
\eea
where $L$ is a parameter of scale transformation and larger than 1. 
{}From Eqs.(\ref{eq:H}) and (\ref{eq:rho_homo}), 
under this scale transformation the variables $H$ and $\rho$ are 
scaled in the following way, 
\bea
   H(t) &\longrightarrow& \L H(t) = LH(Lt), \\
   \rho(t) &\longrightarrow& \L\rho(t) = L^{2}\rho(Lt).
\label{eq:rgrho_homo}
\eea
The equations of motion (\ref{eq:def-H})--(\ref{eq:homo2}) 
are invariant under the scale transformation 
Eqs.(\ref{eq:rgk_homo})--(\ref{eq:rgrho_homo}) 
and equivalently the scaled variables satisfy the original equation.

Second, we define the RG transformation ${\cal R}_L$:
\bea
   {\cal R}_L a(1) = \L a(1), \quad\quad
   {\cal R}_L H(1) = \L H(1), \quad\quad
   {\cal R}_L k(1) = \L k(1).
\label{eq:RG}
\eea

Letting $t=L$, we have formulas 
\bea
   a(t) &=& t^{2/(3\Gamma)}\stackrel{(t)}{a}(1),
\label{eq:reala_Hrho} \\
   H(t) &=& t^{-1}\stackrel{(t)}{H}(1),
\label{eq:realH_Hrho} \\
   k(t) &=& t^{-2(3\Gamma-2)/(3\Gamma)}\stackrel{(t)}{k}(1),
\label{eq:realk}
\eea
which we shall use later to see the long-time behavior of $a$, $H$, and $k$.

Third, we derive the RG equation. 
Letting $L=e^{\tau}$, the infinitesimal transformation of 
$\L a$, $\L H$, and $\L k$ with respect to $\tau$ is  
\bea
  \frac{d\L a}{d\tau} &=& -\frac{2}{3\Gamma} \L a 
  +\frac{\partial\L a}{\partial t},  \nonumber \\
   \frac{d\L H}{d\tau} &=& \L H +\frac{\partial\L H}{\partial t},
\nonumber \\
   \frac{d\L k}{d\tau} &=& \frac{2(3\Gamma-2)}{3\Gamma}\L k 
   +\frac{\partial\L k}{\partial t}.
\label{eq:RGE0} 
\eea

Using the equations of motion (\ref{eq:def-H}), (\ref{eq:H}),
and (\ref{eq:homok}), Eqs.(\ref{eq:RGE0}) can be rewritten as 
\bea
   \frac{d\L a}{d\tau} &=& -\frac{2}{3\Gamma}\L a+\L a\L H,
\nonumber \\
   \frac{d\L H}{d\tau} &=& \L H+\left[
         -\L{H^2}+\frac{2-3\Gamma}{6}\kappa\L\rho \right],
\nonumber \\
   \frac{d\L k}{d\tau} &=& \frac{2(3\Gamma-2)}{3\Gamma}\L{k}.
\label{eq:RG_Hrho}
\eea
These equations (\ref{eq:RG_Hrho}) are the RG equations.

%%%%% fixed point %%%%%%
Here we investigate the fixed point of the RG equations.
The fixed point $( \*{a},\*{H},\*{k} )$ is defined by 
\be
   {\cal R}_L\*{a} = \*{a},\quad\quad
   {\cal R}_L\*{H} = \*{H},\quad\quad
   {\cal R}_L\*{k} = \*{k}.
\label{eq:def_FP}
\ee
The above conditions can be rewritten as 
\be
   \frac{d\L{\*{a}}}{d\tau} = 0,\quad\quad
   \frac{d\L{\*{H}}}{d\tau} = 0,\quad\quad
   \frac{d\L{\*{k}}}{d\tau} = 0.
\label{eq:con_FP}
\ee

{}From Eqs.(\ref{eq:homo2}) and (\ref{eq:RG_Hrho}), the fixed point is 
\be
   \*{a} = \left(\frac{3M\Gamma^2}{4}\right)^{\frac{1}{3\Gamma}},\quad
   \*{H} = \frac{2}{3\Gamma},\quad
   \*{k} = 0,\quad
   \kappa\*{\rho} = 3{\*{H}}^2.
\label{FP_FF}
\ee
This fixed point corresponds to a flat Friedmann universe.

Note that if $\Gamma$ is taken to be $2/3$, there is another 
fixed point where $\*{k}$ is non-zero.
For the non-zero $\*{k}$ case, 
the term of the spatial curvature can be absorbed into 
the term of the energy density of matter because the dependence 
of the scale factor on each term is the same. 
Thus the $\*{k}=0$ case includes the non-zero $\*{k}$ case. 
Hereafter we concentrate on only the $\*{k}=0$ case. 

%%%%%perturbation%%%%%

In order to study the flow in the RG around the fixed point, 
we consider the perturbation around the fixed point.
The perturbed quantities $\L{\delta a}$, $\L{\delta H}$, 
and $\L{\delta k}$ are defined by
\bea
   \L{a} = \*{a}+\L{\delta a}, \quad\quad
   \L{H} = \*{H}+\L{\delta H}, \quad\quad
   \L{k} = \*{k}+\L{\delta k},
\label{eq:def_per}
\eea
where $\L{\delta a}$, $\L{\delta H}$, and $\L{\delta k}$ 
are assumed small qualities.
{}From Eq.(\ref{eq:RG_Hrho}), the perturbed quantities satisfy 
the linearized equations,
\bea
   \frac{d\L{\delta a}}{d\tau} &=& \*{a}\L{\delta H},
\label{eq:per_a_Hrho} \\
   \frac{d\L{\delta H}}{d\tau} &=& -\L{\delta H}
           -\frac{3\Gamma}{2\*{a}}\L{\delta k},
\label{eq:per_H_Hrho} \\
   \frac{d\L{\delta k}}{d\tau} &=& \frac{2(3\Gamma-2)}{3\Gamma}\L{\delta k},
\label{eq:per_k_Hrho} 
\eea
where we neglect the second order term $\delta a^2$ and $\delta H^2$ and 
use the linearized equation of Eq.(\ref{eq:homo2}):
\bea
   \kappa\L{\delta\rho} &=& \frac{4}{\Gamma}\L{\delta H}
    +\frac{3}{\*{a}^2}\L{\delta k}.
\eea

Substituting Eq.(\ref{eq:per_a_Hrho}) into Eq.(\ref{eq:per_H_Hrho}), 
$\L{\delta a}$ satisfies 
\bea
    \frac{d^2\L{\delta a}}{d\tau^2}+\frac{d\L{\delta a}}{d\tau}
    +\frac{3\Gamma}{2\* a}\L{\delta k} &=& 0.
\label{eq:per_FF}
\eea

We solve Eq.(\ref{eq:per_FF});
\bea
   \L{\delta a} &=& f_1 e^{-\tau}+f_2 e^{\frac{2(3\Gamma-2)}{3\Gamma}\tau},
\label{eq:sol_per_FF}
\eea
where $f_1$ and $f_2$ are arbitrary constants.
{}From the solution (\ref{eq:sol_per_FF}), we can see the flow in 
RG around the fixed point.
If $3\Gamma-2 < 0$, this fixed point is an attractor.
On the other hand if $3\Gamma-2 > 0$, there is a single relevant mode.
Note that in the case $3\Gamma-2 > 0$, 
the matter we consider satisfies the strong energy condition.  

{}From the flow in the RG around the fixed point, 
we can see the long-time behavior of the homogeneous and isotropic 
universe.
If $3\Gamma-2 <0$ and setting the initial profile in the vicinity of 
the fixed point $( \* a, \* H, \* k)$, the spacetime will approach to 
the flat Friedmann universe $a(t)={\* a}t^{2/3\Gamma}$. 
On the other hand, if $3\Gamma-2 > 0$, the spacetime will deviate 
from the flat Friedmann universe because there is a relevant mode 
$\delta a(t)=f_2 t^{2(3\Gamma-2)/3\Gamma}$.

In the context of the usual cosmological perturbation 
around a flat Friedmann universe, $f_1$ mode corresponds to 
the decaying mode and $f_2$ mode corresponds to the growing mode, 
which implies the gravitational instability, 
because the matter should satisfy the strong energy condition.

%%%%%%%%%%%%%%%%%%%%%%%%%%%%%%%%%%%%%%%
%     Spherically symmetric case      %
%%%%%%%%%%%%%%%%%%%%%%%%%%%%%%%%%%%%%%%
\subsection{Spherically symmetric case}
\label{sec:sphe}

%%%%% RG equation %%%%%

We consider the spherically symmetric {\it inhomogeneous} case.
In this case, the spatial metric is written by 
\bea
   \gamma_{ij}(x^k,t)dx^idx^j &=& 
         A^2(r,t)dr^2+B^2(r,t)(d\theta^2+\sin^2\theta d\phi^2).
\label{eq:sphemetric}
\eea
Namely, $\grr=A^2(r,t),\gthth=B^2(r,t)$, and 
$\gamma_{\phi\phi}=B^2\sin^2\theta$ while the other components of 
the spatial metric vanish. 
As a simple case, we investigate the universe filled with dust, 
{\it i.e.} $\Gamma=1$ and we can set $u^i=0$ in 
Eqs.(\ref{eq:K}) -- (\ref{eq:ein3}).
The Einstein equations are 
\bea
   \dot{\gamma_{ij}} &=& 2K_{ij}
\label{eq:sphe0} \\
   \dot{K_{ij}} &=& -\3R_{ij}+2K_{il}K^l_j-KK_{ij}
        +\frac{1}{2}\kappa\rho\gamma_{ij},
\label{eq:sphe1} \\
   \kappa\rho &=& \frac{1}{2}\left[\3R+K^2-K^m_lK^l_m\right],
\label{eq:sphe2} \\
   K^j_{i;j}-K_{,i} &=& 0.
\label{eq:sphe3}
\eea

Here we consider the following scale transformation.
\bea
   r &\longrightarrow& Lr, \nonumber 
\nonumber \\
   t   &\longrightarrow& L^{\alpha}t, 
\nonumber \\
   \grr(r,t) &\longrightarrow& 
   \L{\grr}(r,t)\equiv L^{2-2\alpha}\grr (Lr,L^{\alpha}t),
\label{eq:rgrr} \\
   \gthth(r,t) &\longrightarrow& 
   \L{\gthth}(r,t)\equiv L^{-2\alpha}\gthth (Lr,L^{\alpha}t),
\label{eq:rgthth} \\
   \Krr(r,t) &\longrightarrow& 
   \L\Krr(r,t)\equiv L^{2-\alpha}\Krr (Lr,L^{\alpha}t),
\label{eq:rgKrr} \\
   \Kthth(r,t) &\longrightarrow& 
   \L\Kthth(r,t)\equiv L^{-\alpha}\Kthth (Lr,L^{\alpha}t),
\label{eq:rgKthth} \\
   \rho(r,t) &\longrightarrow& 
   \L\rho(r,t)\equiv L^{2\alpha}\rho (Lr,L^{\alpha}t),
\label{eq:rgrho}
\eea
where $\alpha$ is an arbitrary constant because the coordinate transformation,
$r\rightarrow r^\beta$, yields substitution of $\alpha/\beta$ 
for $\alpha$ in Eqs.(\ref{eq:rgrr})--(\ref{eq:rgrho}).
Without loss of generality, we take $\alpha$ to be positive.
Because of the scale invariance of the Einstein equations 
Eqs.(\ref{eq:sphe0}) -- (\ref{eq:sphe3}), the scaled variables 
$\L{\grr}$, $\L{\gthth}$, $\L{\Krr}$, $\L{\Kthth}$, 
and $\L{\rho}$ also satisfy Eqs.(\ref{eq:sphe0}) -- (\ref{eq:sphe3}). 

We derive the RG equation. 
Letting $L=e^{\tau}$, the infinitesimal transformation of 
$\L{\g{ij}}$ and $\L{\K{ij}}$ with respect to $\tau$ is  
\bea
   \frac{d\L\grr}{d\tau} &=& 2(1-\alpha)\L\grr+r\partial_r\L\grr+2\alpha\L\Krr,
\label{eq:RGgrr} \\
   \frac{d\L\gthth}{d\tau} &=& 
   -2\alpha\L\gthth+r\partial_r\L\gthth+2\alpha\L\Kthth,
\label{eq:RGgthth} \\
   \frac{d\L\Krr}{d\tau} &=& (2-\alpha)\L\Krr+r\partial_r\L\Krr
        +\alpha\left[
        \frac{1}{4}\L{\3 R}\L\grr-\L{\3 R_{rr}}+2\L{K_{rl}}\L{K_r^l}
        -\L K\L\Krr
        +\frac{1}{4}\left(\L{K^2}\L\grr-\L{K_l^m}\L{K_m^l}\L\grr 
        \right )\right],
\label{eq:RGKrr} \\
   \frac{d\L\Kthth}{d\tau} &=& -\alpha\L\Kthth+r\partial_r\L\Kthth
        +\alpha\left[
        \frac{1}{4}\L{\3 R}\L\gthth-\L{\3 R_{\theta\theta}}
        +2\L{K_{\theta l}}\L{K_{\theta}^l}-\L K\L\Kthth
        +\frac{1}{4}\left(\L{K^2}\L\gthth-\L{K_l^m}\L{K_m^l}\L\gthth 
        \right )\right],
\label{eq:RGKthth}
\eea
where $\L{\3 R_{ij}}$ is the Ricci tensor associated with $\L{\gamma_{ij}}$.
In the derivation of  Eqs.(\ref{eq:RGgrr}) -- (\ref{eq:RGKthth}), 
the equations of motion (\ref{eq:sphe0}) and (\ref{eq:sphe1}) are used.
These equations (\ref{eq:RGgrr}) -- (\ref{eq:RGKthth}) are the RG equations.

In terms of $\L{A}$ and $\L{B}$ 
($\L{A^2}=\L{\grr}$ and $\L{B^2}=\L{\gthth}$),
the RG equations (\ref{eq:RGgrr})--(\ref{eq:RGKthth}) read 
\bea
&& \frac{d^2\L{A}}{d\tau^2}
   +\left[2(\alpha-1)+\frac{1}{\L{B}}\frac{d\L{B}}{d\tau}
   -\frac{r\L{B'}}{\L{B}}\right]\frac{d\L{A}}{d\tau}-2r\frac{d\L{A'}}{d\tau} =
\nonumber \\ &&
   -r^2\L{A''}+(2\alpha-3)r\L{A'}
   -\frac{r\L{B'}}{\L B}\L{A}
   +\frac{1}{2}\left(\frac{r\L{B'}}{\L{B}}\right)^2\L{A}
   -\frac{r^2\L{A'}\L{B'}}{\L{B}}
   -\frac{\alpha^2-4\alpha+2}{2}\L{A} 
\nonumber \\ &&
   +\frac{\alpha^2(\L{A^3}-2\L{A'}\L{B}\L{B'}
   -\L{A}\L{B'^2}+2\L{A}\L{B}\L{B''})}{2\L{A^2}\L{B^2}}
%\nonumber \\ &&
   +\frac{1}{\L{B^2}}\left[
   \frac{\L{A}}{2}\frac{d\L{B}}{d\tau}+\L{A}\L{B}-r\L{A}\L{B'}+r\L{A'}\L{B}
   \right]\frac{d\L{B}}{d\tau},
\label{eq:RG-A} \\
&& \frac{d^2\L{B}}{d\tau^2}
   +\left[2\alpha+\frac{1}{2\L{B}}\frac{d\L{B}}{d\tau}
   -\frac{r\L{B'}}{\L{B}}\right]\frac{d\L{B}}{d\tau}
   -2r\frac{d\L{B'}}{d\tau} =    
\nonumber \\ &&
   -r^2\L{B''}+(2\alpha-1)r\L{B'}
   -\frac{r^2\L{B'^2}}{2\L{B}}
   +\frac{\alpha^2(\L{B'^2}-\L{A^2})}{2\L{A^2}\L{B}}
   -\frac{\alpha^2}{2}\L{B},
\label{eq:RG-B}
\eea
where the prime denotes the derivative with respect to $r$. 
From Eqs.(\ref{eq:sphe2}) and (\ref{eq:sphe3}), we obtain 
\bea
&& \kappa\L{\rho} = \frac{2}{\alpha^2\L{A}}\left[\alpha
   +\frac{1}{\L{B}}\frac{d\L{B}}{d\tau}-\frac{r\L{B'}}{\L{B}}\right]
   \frac{d\L{A}}{d\tau}
   +\frac{1}{\alpha^2\L{B}}\left[2(2\alpha-1)
   -2\frac{r\L{A'}}{\L{A}}-2\frac{r\L{B'}}{\L{B}}
   +\frac{1}{\L{B}}\frac{d\L{B}}{d\tau}\right]\frac{d\L{B}}{d\tau}
\nonumber \\ &&
   +\frac{3\alpha-2}{\alpha}
   -\frac{2}{\alpha}\frac{r\L{A'}}{\L{A}}
   -\frac{2(2\alpha-1)}{\alpha^2}\frac{r\L{B'}}{\L{B}}
   +\frac{2}{\alpha^2}\frac{r^2\L{A'}\L{B'}}{\L{A}\L{B}}
   +\frac{1}{\alpha^2}\frac{r^2\L{B'^2}}{\L{B^2}}
\nonumber \\ &&
   +\frac{1}{\L{A^3}\L{B^2}}\left[
   \L{A^3}+2\L{A'}\L{B}\L{B'}-\L{A}\L{B'^2}-2\L{A}\L{B}\L{B''}\right],
\label{eq:def-rho} \\
&& \L{B'}\frac{d\L{A}}{d\tau}-\L{A}\frac{d\L{B'}}{d\tau}
   -r\L{A'}\L{B'}+r\L{A}\L{B''} = 0.
\eea

Letting $t=L^{\alpha}$, the original variables $A(r,t)$ and $B(r,t)$ 
are expressed by the scaled variables $\L{A}$ and $\L{B}$:
\bea
   A(r,t) &=& t^{(\alpha-1)/\alpha}\stackrel{(t^{1/\alpha})}
   A(rt^{-1/\alpha},1),
\label{eq:rA}\\
   B(r,t) &=& t\stackrel{(t^{1/\alpha})}B(rt^{-1/\alpha},1).
\label{eq:rB}
\eea

%%%%% fixed point %%%%%%
Here we investigate the fixed point of the RG equations (\ref{eq:RG-A}) 
and (\ref{eq:RG-B}) defined by 
\bea
   \frac{d\*{A}}{d\tau}=0,&& \quad\quad
   \frac{d\*{B}}{d\tau}=0.
\label{eq:fp-AB}
\eea
At the fixed point, 
\be
   A(r,t) = t^{(\alpha-1)/\alpha}\stackrel{(t^{1/\alpha})}
   {\* A}(rt^{-1/\alpha},1) = 
   t^{(\alpha-1)/\alpha}\times\mbox{( function of $rt^{-1/\alpha}$ only )}
\label{eq:fp-A}
\ee
and 
\be
   B(r,t) = t\stackrel{(t^{1/\alpha})}{\* B}(rt^{-1/\alpha},1) = 
   t\times\mbox{( function of $rt^{-1/\alpha}$ only )}
\label{eq:fp-B} 
\ee
is a self-similar solution.

In the spherically symmetric spacetime filled with dust, 
the general solution of the Einstein equations is the Tolman-Bondi 
solution (\ref{eq:TB-A}) -- (\ref{eq:TB-B}) in the Appendix.
Therefore we can obtain the fixed point from the Tolman-Bondi solution 
with self-similarity rather than solving the equations (\ref{eq:fp-AB})
 directly.
The precise form of these are 

For $c=0$ :
\bea
   \*{A}(r,1) &=& \frac{\alpha r^{\alpha/3-1}(1-3pr^{\alpha})}
               {3(1-pr^{\alpha})^{\frac{1}{3}}},
\\
   \*{B}(r,1) &=& r^{\alpha/3}(1-pr^{\alpha})^{\frac{2}{3}},
\\
   \kappa\*\rho(r,1) &=& \frac{4}{3(1-pr^{\alpha})(1-3pr^{\alpha})},
\label{eq:FP-TB-c0}
\eea

For $c > 0$ :
\bea
   \*{A}(r,1) &=& \frac{\alpha}{(1+c)^{1/2}r}\left[
               \frac{2}{9c}(\cosh\eta-1)r^{\alpha}
               -c^{1/2}\frac{\sinh\eta}{\cosh\eta-1}\right],
\\
   \*{B}(r,1) &=& \frac{2}{9c}r^{\alpha}(\cosh\eta-1),
\\
   \sinh\eta-\eta &=& \frac{9c^{3/2}}{2}(r^{-\alpha}-p),
\\
   \kappa\*\rho(r,1) &=& \frac{9c^2}{r^{\alpha}(\cosh\eta-1)^2\left[
   \frac{2r^{\alpha}}{9c}(\cosh\eta-1)
   -c^{1/2}\frac{\sinh\eta}{\cosh\eta-1}\right]},
\label{eq:FP-TB-c+}
\eea

For $c < 0$ :
\bea
   \*{A}(r,1) &=& \frac{\alpha}{(1-|c|)^{1/2}r}\left[
               \frac{2}{9|c|}(1-\cos\eta)r^{\alpha}
               -|c|^{1/2}\frac{\sin\eta}{1-\cos\eta}\right],
\\
   \*{B}(r,1) &=& \frac{2}{9|c|}r^{\alpha}(1-\cos\eta),
\\
   \eta-\sin\eta &=& \frac{9|c|^{3/2}}{2}(r^{-\alpha}-p),
\\
   \kappa\*\rho(r,1) &=& \frac{9c^2}{r^{\alpha}(1-\cos\eta)^2\left[
   \frac{2r^{\alpha}}{9|c|}(1-\cos\eta)
   -|c|^{1/2}\frac{\sin\eta}{1-\cos\eta}\right]},
\label{eq:FP-TB-c-}
\eea
where $c$ and $p$ are constants.

The constant $c$ can be interpreted as the total energy of the universe 
in the analogy of the Newtonian mechanics.
By the signature of the constant $c$, these fixed points are classified 
into the following three. 
The universe with $c=0$ is similar to the flat Friedmann universe.
The universes with a positive $c$ or a negative $c$ are similar to 
the open and closed Friedmann universes, respectively.
Especially when $c=p=0$, the above fixed point coincides with the flat 
Friedmann universe and the spacetime becomes homogeneous.

In the context of the RG, we can treat the time evolution of the field 
variables as the map from a set of initial data to another. 
If the initial data is taken to be the above fixed point $\*{A}$, $\*{B}$, 
and $\*{\rho}$, the spacetime will evolve into the Tolman-Bondi solution 
with self-similarity. 
In the cases of $c=0$ and $c>0$, the fixed point is not regular if $p>0$.
For $c<0$, the fixed point has singularities irrespective of the signature 
of $p$ because the spacetime is similar to the closed Friedmann universe 
and will recollapse. 
Since we should set a regular initial data in the physical situation, 
we investigate only the case of $c=0$ ($p<0$) and $c>0$ ($p\le 0$) 
where the fixed point is everywhere regular.
Note we exclude the case of $c=p=0$ 
because we have already studied it in the previous subsection.

%%%%%perturbation%%%%%
To study the behavior of the flow in the RG around the fixed 
point, we consider the linear perturbation around the fixed point.
The linear perturbation around the self-similar Tolman-Bondi solution 
has also been discussed by Tomita \cite{Tomita} 
in a different context from ours \cite{CT}. 
For simplicity, we concentrate on the spherical modes of 
linear perturbation.
The perturbed quantities $\L{\delta A}$ and $\L{\delta B}$ are 
defined by 
\bea
   \L\grr &=& \*{A}^2+2\* A\L{\delta A},
\nonumber \\
   \L\gthth &=& \*{B}^2+2\* B\L{\delta B}.
\label{eq:per}
\eea

We assume the spatial metric variables for $\delta\L A$ and $\delta\L B$ 
in the following form, 
\bea
   \L{\delta A} &=& a(r)e^{\omega\tau},
\nonumber \\
   \L{\delta B} &=& b(r)e^{\omega\tau}.
\label{eq:sepa}
\eea
The perturbed quantities $a(r)$ and $\delta\L{\rho}$ are expressed by $b(r)$;
\bea
\frac{a}{\*{A}} &=& \frac{b'}{\*{B'}}-\frac{c_1}{2(1+c)}r^\omega,
\label{eq:per-a} \\
\frac{\delta{\L\rho}}{{\*\rho}} &=& e^{\omega\tau}\left[
      \frac{9(\omega+\alpha)}{4\alpha}c_2r^\omega
      -2\frac{b}{\*{B}}-\frac{b'}{\*{B'}}\right],
\label{eq:per-rho}
\eea
where $c_1$ and $c_2$ are arbitrary constants ({\it see Appendix}). 

As for the spherical modes of the perturbation, we can easily obtain 
the solutions.

For $c=0$: 

\bea
b(r) &=& r^{\omega}\left[\frac{9}{20}c_1r^{-\alpha/3}(1-pr^{\alpha})^{4/3}
         +\frac{3}{4}c_2r^{\alpha/3}(1-pr^{\alpha})^{2/3}
         -\frac{2}{3}c_3r^{4\alpha/3}(1-pr^{\alpha})^{-1/3}\right],
\label{eq:b0}
\eea
where $c_3$ is another arbitrary constant.
The density contrast is 
\bea
\frac{\delta\L{\rho}}{\*{\rho}} &=& e^{\omega\tau}r^{\omega}\left\{
      -\frac{9}{20\alpha}c_1r^{-2\alpha/3}(1-pr^\alpha)^{2/3}
      (1-3pr^{\alpha})^{-1}
      \left[3\omega+\alpha-3(\omega+3\alpha)pr^{\alpha}\right]
\right. \nonumber \\ &&\left.
      -\frac{9\omega}{2\alpha}c_2pr^\alpha(1-3pr^\alpha)^{-1}
      +\frac{2}{\alpha}c_3r^\alpha(1-pr^\alpha)^{-1}(1-3pr^\alpha)^{-1}
      \left[\omega+2\alpha-(\omega+3\alpha)pr^\alpha\right]\right\}.
\label{eq:per-rho0} 
\eea
In the expression for the linear perturbation Eq.(\ref{eq:per-rho0}) 
there are three terms corresponding to $c_1$,$c_2$, and $c_3$ 
so that there should be a gauge mode hidden in Eq.(\ref{eq:per-rho0}) 
because the number of physical modes has to be two. 
Actually there remains a gauge freedom corresponding to the coordinate 
transformation of $r$ (\ref{eq:ctran}).
The gauge mode is given by 
\bea
\frac{\delta\L{\rho}_g}{\*{\rho}} &=& 
  fr^{\omega+1}e^{\omega\tau}\frac{\*{\rho'}}{\*{\rho}}
\nonumber \\
   &=&e^{\omega\tau}r^\omega f\left[
   2\alpha pr^\alpha(2-3pr^\alpha)(1-pr^\alpha)^{-1}(1-3pr^\alpha)^{-1}\right],
\label{eq:per-rho0g}
\eea
with $f$ being an arbitrary constant.
Because we should fix the freedom of gauge we choose $f$ so that 
$(\delta\L{\rho}+\delta\L{\rho}_g)/\*{\rho}$ behave 
as nicely as possible at $r=\infty$ because we are interested 
in the perturbation modes which are finite at $r=\infty$. 

We use the following condition as a convenient gauge condition. 
\bea
 f &=& -\frac{9(\omega+3\alpha)}{40\alpha^2}p^{2/3}c_1
       +\frac{3\omega}{4\alpha^2}c_2-\frac{\omega+3\alpha}{3\alpha^2p}c_3.
\label{eq:gauge-c0}
\eea

We fix the gauge mode by the above condition and obtain 
the physical perturbation 
$\delta\L{\tilde{\rho}}=\delta\L{\rho}+\delta\L{\rho}_g$ as follows.
\bea
 \frac{\delta\L{\tilde{\rho}}}{\*{\rho}} &=& e^{\omega\tau}r^\omega\left\{
  \Delta_1(1-3pr^\alpha)^{-1}\left[
  p^{-2/3}r^{-2\alpha/3}(1-pr^\alpha)^{2/3}
  \left[3\omega+\alpha-3(\omega+3\alpha)pr^\alpha\right]
\right.\right. \nonumber \\ && \left.\left.
  +(\omega+3\alpha)pr^\alpha(2-3pr^\alpha)(1-pr^\alpha)^{-1}\right]
  +\Delta_2pr^\alpha(1-pr^\alpha)^{-1}(1-3pr^\alpha)^{-1}\right\},
  \label{eq:per-c0}
\eea
where
\bea
 \Delta_1 &=& -\frac{9p^{2/3}}{20\alpha}c_1,\\
 \Delta_2 &=& -\frac{3\omega}{2\alpha}(c_2-\frac{4}{9p}c_3).
\eea

\begin{figure} 
\begin{center}
  \leavevmode
  \epsfbox{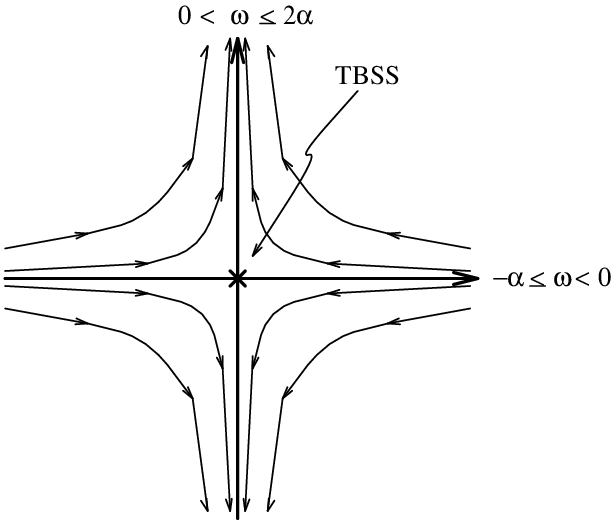}
\caption{
TBSS represents the Tolman-Bondi solution with self-similarity 
in the case of $c=0$ and $p<0$. 
The axes correspond to the modes of linear perturbation 
Eq.(\ref{eq:per-c0}).
}
\label{fig:c0}
\end{center}
\end{figure}

Since we consider only the case of $p<0$, 
the coordinate $r$ can be taken from $0$ to $\infty$. 
We demand the regularity condition that $\delta\L{\tilde{\rho}}/\*{\rho}$ 
should be finite at the boundary, $r=0$ and $r=\infty$. 
This condition implies that $\Delta_1$ modes with 
$2\alpha/3\le\omega\le\alpha$ and 
$\Delta_2$ modes with $-\alpha\le\omega\le\alpha$ are allowed.
The mode with $\omega=0$ corresponds to change of $p$ in the self-similar 
solution Eqs.(\ref{eq:FP-TB-c0}) and $\*{\rho}$ remains constant 
independent of $\tau$ in the direction.
Although this fixed point is not a repeller, 
it has many relevant modes, 
$\Delta_1$ with $0<\omega\le\alpha$ and $\Delta_2$ with 
$0<\omega\le\alpha$. 
Note that a suitable linear combination of the $\Delta_1$ and $\Delta_2$ modes 
will have an asymptotic behavior $\approx r^{\omega-2\alpha}$ at $r=\infty$. 
For such modes, $2\alpha/3\le\omega\le 2\alpha$ is allowed.
These modes which satisfy the regularity condition are not discrete 
but continuous.
We suppose that this special feature arises because our matter 
is assumed to be dust.
To summarize, the possible value of $\omega$ ranges from 
$-\alpha$ to $2\alpha$.
The flow of RG in the vicinity of the fixed point is 
shown in Fig.(\ref{fig:c0}).

In the case of $c=p=0$ where the fixed point corresponds to the flat Friedmann 
universe, the $c_2$ modes in the density contrast (\ref{eq:per-rho0})
are gauge modes.
The regularity condition implies that $c_1$ mode with $\omega=2\alpha/3$ 
and $c_3$ mode with $\omega=-\alpha$ are allowed.
This result corresponds to the homogeneous and isotropic case in the previous 
section.
Compared with the usual cosmological perturbation in the synchronous 
comoving reference frame, this result appears to be strange because 
the only spherical modes of the linear perturbation allowed are 
constant in space.
However, the time coordinate used in this RG method is different from 
the usual cosmic time coordinate, the solution allowed by the regularity 
condition in each case does not coincide in general.
Moreover, in the homogeneous universe, there is no non-trivial characteristic 
profile of field variables. 
If the fixed point is a homogeneous universe, 
the RG method may have no advantage 
since the RG approach respects the self-similar profile.
But if the fixed point is an inhomogeneous universe, 
we believe that the RG method may be useful.

Compared with $c=0(p<0)$ case, the value of $\omega$ allowed in the case 
$(p=0)$ is the lower limit in the case $(p<0)$.
The effect of nonlinearity of gravity makes 
the growth rate of the density contrast large.

For $c > 0$:

\bea
b(r) &=& r^{\omega}\left\{-\frac{1}{9c^2}c_1
      \left[2(\cosh\eta-1)-\frac{3\sinh\eta(\sinh\eta-\eta)}{\cosh\eta-1}
      \right]
\right. \nonumber \\ && \left.
      +\frac{1}{2c}c_2
      \left[\cosh\eta-1-\frac{\sinh\eta(\sinh\eta-\eta)}{\cosh\eta-1}\right]
      -\frac{c^{1/2}\sinh\eta}{\cosh\eta-1}c_3\right\}.
\label{eq:b+}
\eea
The density contrast is 
\bea
&&\frac{\delta\L{\rho}}{\*{\rho}} = \frac{e^{\omega\tau}r^{\omega}}
      {\alpha\left[\frac{2r^\alpha}{9c}(\cosh\eta-1)
      -\frac{c^{1/2}\sinh\eta}{\cosh\eta-1}\right]}\left\{
\right. \nonumber \\ && \left.
      \frac{1}{9c^2}c_1
      \left[(\omega+3\alpha)r^\alpha\left[
      2(\cosh\eta-1)-\frac{3\sinh\eta(\sinh\eta-\eta)}{\cosh\eta-1}\right]
%%%%%
%\right.\right. \nonumber \\ && \left.\left.
%%%%%   
      +\frac{27c^{3/2}\alpha}{2}\left[-\frac{\sinh\eta}{\cosh\eta-1}
      +\frac{(2\cosh\eta+1)(\sinh\eta-\eta)}{(\cosh\eta-1)^2}\right]\right]
\right. \nonumber \\ && \left.
      -\frac{1}{2c}c_2
      \left[r^\alpha\left[2\alpha(\cosh\eta-1)
      -(\omega+3\alpha)\frac{\sinh\eta(\sinh\eta-\eta)}{\cosh\eta-1}\right]
%%%%%
%\right.\right. \nonumber \\ && \left.\left.
%%%%%   
      +\frac{9c^{2/3}}{2}\left[\frac{(\omega-\alpha)\sinh\eta}{\cosh\eta-1}
      +\frac{\alpha(2\cosh\eta+1)(\sinh\eta-\eta)}{(\cosh\eta-1)^2}
      \right]\right]
\right. \nonumber \\ && \left.
      +c^{1/2}c_3\left[(\omega+3\alpha)r^\alpha\frac{\sinh\eta}{\cosh\eta-1}
      -\frac{9\alpha c^{2/3}(2\cosh\eta+1)}{2(\cosh\eta-1)^2}\right]\right\}.
\label{eq:per-rho+}
\eea
The gauge mode is given by 
\bea
&&\frac{\delta\L{\rho}_g}{\*{\rho}} = 
      fr^{\omega+1}e^{\omega\tau}\frac{\*{\rho'}}{\*{\rho}}
\nonumber \\
      &&= -\frac{\alpha fe^{\omega\tau}r^{\omega}}
      {\left[\frac{2r^\alpha}{9c}(\cosh\eta-1)
      -\frac{c^{1/2}\sinh\eta}{\cosh\eta-1}\right]}\left[
      \frac{4}{9c}r^\alpha(\cosh\eta-1)-\frac{4c^{1/2}\sinh\eta}{\cosh\eta-1}
      +\frac{9c^2r^{-\alpha}(2\cosh\eta+1)}{2(\cosh\eta-1)^2}\right].
\label{eq:per-rhog+}
\eea

Similarly to the case of $c=0$, we demand the regularity condition 
at $r=0$ and $r=\infty$. 

\begin{figure} 
\begin{center}
  \leavevmode
  \epsfbox{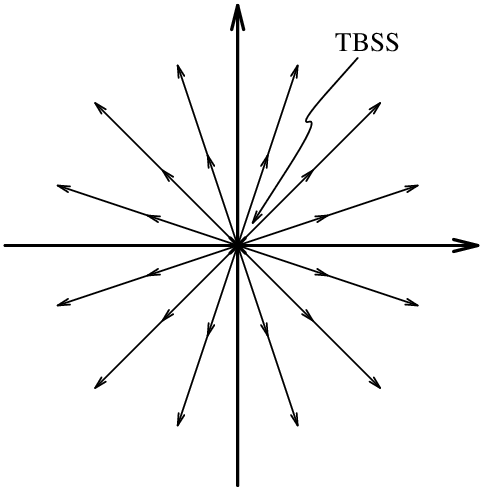}
\caption{
TBSS represents the Tolman-Bondi solution with self-similarity 
in the case of $c>0$ and $p<0$. 
The axes correspond to the modes of linear perturbation 
Eq.(\ref{eq:per-c+}).
}
\label{fig:c+}
\end{center}
\end{figure}

\begin{enumerate}

\item case 1 ($p < 0$):

We use the following convenient gauge condition,
\bea
&& f = \frac{1}{4\alpha^2}\left\{
       \frac{\omega+3\alpha}{c}c_1\left[
       2-3\frac{\sinh\eta_0(\sinh\eta_0-\eta_0)}{(\cosh\eta_0-1)^2}\right]
\right. \nonumber \\ && \left.
       -\frac{9}{2}c_2\left[2\alpha-(\omega+3\alpha)
       \frac{\sinh\eta_0(\sinh\eta_0-\eta_0)}{(\cosh\eta_0-1)^2}\right]
       +\frac{9c^{2/3}(\omega+3\alpha)(\sinh\eta_0)}
             {(\cosh\eta_0-1)^2}c_3\right\},
\label{eq:gauge-c+}
\eea
where $\eta=\eta_0$ corresponds to $r\rightarrow\infty$ and $\eta_0$ 
is thus given by $\sinh\eta_0-\eta_0=-9pc^{3/2}/2$. 

By using the above condition Eq.(\ref{eq:gauge-c+}), 
we obtain the physical perturbation 
$\delta\L{\tilde{\rho}}=\delta\L{\rho}+\delta\L{\rho}_g$ as follows. 
\bea
&&\frac{\delta\L{\tilde{\rho}}}{\*{\rho}} = 
      \frac{e^{\omega\tau}r^{\omega}}
      {\alpha\left[\frac{2r^\alpha}{9c}(\cosh\eta-1)
      -\frac{c^{1/2}\sinh\eta}{\cosh\eta-1}\right]}\left\{
\right. \nonumber \\ && \left.
      \Delta_1^{+}
      \left[3(\omega+3\alpha)r^\alpha(\cosh\eta-1)\left[
      \frac{\sinh\eta_0(\sinh\eta_0-\eta_0)}{(\cosh\eta_0-1)^2}
      -\frac{\sinh\eta(\sinh\eta-\eta)}{(\cosh\eta-1)^2}\right]
\right.\right. \nonumber \\ && \left.\left.
      +\frac{9c^{3/2}}{2}\left[\left[4\omega+9\alpha-6(\omega+3\alpha)
      \frac{\sinh\eta_0(\sinh\eta_0-\eta_0)}{(\cosh\eta_0-1)^2}
      \right]\frac{\sinh\eta}{\cosh\eta-1}
%%%%%
%\right.\right.\right. \nonumber \\ && \left.\left.\left.
%%%%%   
      +3\alpha\frac{(2\cosh\eta+1)(\sinh\eta-\eta)}{(\cosh\eta-1)^2}
      \right]\right]
\right. \nonumber \\ && \left.
      +\Delta_2^+\left[
      (\omega+3\alpha)r^\alpha(\cosh\eta-1)\left[
      \frac{\sinh\eta}{(\cosh\eta-1)^2}-\frac{\sinh\eta_0}{(\cosh\eta_0-1)^2}
      \right]-\frac{9c^{3/2}\alpha(2\cosh\eta+1)}{2(\cosh\eta-1)^2}
\right.\right. \nonumber \\ && \left.\left.
      +\frac{9(\omega+3\alpha)c^{2/3}\sinh\eta_0}{4(\cosh\eta_0-1)^2}
      \left[\frac{4\sinh\eta}{\cosh\eta-1}
      \frac{9c^{3/2}(2\cosh\eta+1)}{2(\cosh\eta-1)^2r^\alpha}\right]
      \right]\right\},
\label{eq:per-c+}
\eea
where
\bea
 \Delta_1^+ &=& \frac{1}{9c^2}c_1, \\
 \Delta_2^+ &=& -\frac{9pc^{1/2}}{4}(c_2-\frac{4}{9p}c_3), \\
\eea
and $\sinh\eta-\eta=9c^{3/2}(r^{-\alpha}-p)/2$. 

The regularity condition at $r=0$ and $r=\infty$ implies that
$\Delta_1^+$ modes with $\omega=\alpha$ and 
$\Delta_2^+$ modes with $0\le\omega\le\alpha$ are allowed. 
In this case, this fixed point is a repeller up to the zero mode 
because all other modes which satisfy the regularity condition 
have a positive $\omega$. 
Note that a suitable linear combination of the $\Delta_1^+$ and $\Delta_2^+$ 
modes will have an asymptotic behavior $\approx r^{\omega-2\alpha}$. 
Therefore the possible value of $\omega$ ranges from $0$ to $2\alpha$. 
The flow of RG in the vicinity of the fixed point is 
shown in Fig.(\ref{fig:c+}).

\item case 2 ($p=0$):

The gauge condition which we use is,
\bea
 f &=& -\frac{9}{4\alpha}c_2. 
\label{eq:gauge-c+p0}
\eea

By using the above condition Eq.(\ref{eq:gauge-c+p0}), 
we obtain the physical perturbation,
\bea
&&\frac{\delta\L{\tilde{\rho}}}{\*{\rho}}= \frac{e^{\omega\tau}r^{\omega}}
      {\alpha\left[\frac{2r^\alpha}{9c}(\cosh\eta-1)
      -\frac{c^{1/2}\sinh\eta}{\cosh\eta-1}\right]}\left\{
%%%%%
%\right. \nonumber \\ && \left.
%%%%%
      \Delta_1^+
      \left[(\omega+3\alpha)r^\alpha\left[
      2(\cosh\eta-1)-\frac{3\sinh\eta(\sinh\eta-\eta)}{\cosh\eta-1}\right]
\right.\right. \nonumber \\ && \left.\left.
      +\frac{27c^{3/2}\alpha}{2}\left[-\frac{\sinh\eta}{\cosh\eta-1}
      +\frac{(2\cosh\eta+1)(\sinh\eta-\eta)}{(\cosh\eta-1)^2}\right]\right]
\right. \nonumber \\ && \left.
      +\Delta_2^+\left[\frac{(\omega+3\alpha)r^\alpha\sinh\eta}{\cosh\eta-1}
      -\frac{9\alpha c^{2/3}(2\cosh\eta+1)}{2(\cosh\eta-1)^2}\right]\right\},
 \label{eq:per-c+p0}
\eea
where
\bea
 \Delta_1^+ &=& \frac{1}{9c^2}c_1, \\
 \Delta_2^+ &=& c^{1/2}c_3.
\eea
\end{enumerate}

The regularity condition at $r=0$ and $r=\infty$ implies that
no modes are allowed.

{}From the linear perturbation analysis around the fixed point, 
we see that the long-time behavior of the spherically symmetric 
dust universe is separated into two types.
One is the case that the fixed point is a repeller.
In this case, the Tolman-Bondi solution with self-similarity dose not 
play an important role in expanding universe because this fixed point 
is unstable and the spacetime will diverge from this fixed point.
In the other case, the fixed point has both relevant and irrelevant modes.
Although this fixed point is not a repeller, 
it has continuously many relevant modes.
Thus it is not as straightforward as in the case of gravitational 
collapse \cite{HKA} to extract the long-time behavior of the universe, 
because it is sensitive to the initial condition and therefore we cannot 
uniquely predict the outcome. 
In the final section, we briefly discuss how to treat the fixed point 
which has many relevant modes of the perturbation. 

%%%%%%%%%%%%%%%%%%%%%%%%%%%%%%%%%%%%%%%
%      Summary and Discussions        %
%%%%%%%%%%%%%%%%%%%%%%%%%%%%%%%%%%%%%%%
\section{Summary and Discussions}
\label{sec:dis}

We considered the spherically symmetric but inhomogeneous universe 
filled with dust, where the Einstein equations have scale 
invariance Eqs.(\ref{eq:rgrr})--(\ref{eq:rgKthth}) and applied 
the renormalization group method to study its long-time asymptotics. 
The fixed point of the RG transformation is a self-similar 
solution with scale invariance of the Einstein equations.
In order to study the flow of the RG around this fixed point, 
the linear perturbation analysis is used.
We impose the perturbation on the regularity at the boundary 
where the radial coordinate $r$ equals zero or infinity. 
This boundary means that the area radius equals zero or infinity 
in the case of $c=0$, on the other hand, in the case of $c>0$, 
it equals a finite or infinity. 
The fixed point is the Tolman-Bondi solution 
with self-similarity, which includes the flat Friedmann universe.
The behavior of the fixed point is separated into two types. 
Both types have many relevant modes of the perturbation.

The Tolman-Bondi solution with self-similarity 
is unstable against almost all spherical modes of linear perturbation. 
The spacetime will deviate from this fixed point.
It is necessary to study the non-spherical mode of perturbation 
to say something more definite. 
In the cosmological problem, only the statistical 
quantities are meaningful if we think of comparison with observations.
There are some works in the RG approach \cite{PGHL,BDGP,CP} 
on the universe which has a hierarchical structure.
We may contemplate further development of the RG approach to cosmology 
formulated in the present work by introducing some kind of volume or 
statistical average for observables like energy density and the Hubble 
constant of the universe. 
The statistical concept is needed not only for comparison with 
observations but also for us to proceed further in the analysis of the RG 
equation because we have continuously many relevant (growing) modes around 
the fixed points. That is, the long-time behavior of the universe is 
sensitive to the initial configuration which we have no a priori control 
and we have to consider statistical likelihood of the initial values. 

We remark that the introduction of volume average in a finite region of the 
universe potentially introduces the scale invariance violation 
by hand because the exact scale invariance holds only for an infinite space. 
Note that in quantum field theories and statistical physics of 
the second order phase transition the scale invariance violations 
are hidden in the form of cut-off of the spectrum of physical modes. 
We shall elaborate our present observation in our future work. 

The self-similar solution given by Eqs.(\ref{eq:rA}) and 
(\ref{eq:rB}) through the fixed point of the RG equation 
is essentially a function of $t^{-1/\alpha}r=t^{1-1/\alpha}r/t$, 
which is roughly the fraction of physical distance to the horizon scale 
of the Friedmann universe. 
Also note that in the case of non-linear diffusion equation 
Eq.(\ref{eq:ru}) implies the self-similar solution is a function of 
the ratio of the distance $x$ to the diffusion length $\sqrt{t}$. 
In the both cases, the self-similar solution is a function of 
the distance in the unit of physically relevant time dependent scale. 
We believe this is a general phenomenon and the physical background of 
the RG equation which governs how dynamical variables deviate from 
the self-similar solution.

%%%%%%%%%%%%%%%%%%%%%%%%%%%%%%%%%%%%%%%
%           Acknowledgments          %
%%%%%%%%%%%%%%%%%%%%%%%%%%%%%%%%%%%%%%%
\section*{Acknowledgments}
O. I. would like to thank Professor H.Ishihara and K.Nakamura 
for discussions.
The research is supported in part by Japan Society for the Promotion of
Science (O. I.).
This work is partially supported by grant-in-aid by the Ministry 
of Education, Science, Sports, and Culture of Japan 
(A. H., 09640341 and T. K., 012-10096097)

%%%%%%%%%%%%%%%%%%%%%%%%%%%%%%%%%%%%%%%
%             Appendix                %
%%%%%%%%%%%%%%%%%%%%%%%%%%%%%%%%%%%%%%%
\begin{appendix}
\section{Tolman-Bondi solution with self-similarity}

In the spherically symmetric universe filled with dust, 
the most general solution of the Einstein equations is the Tolman-Bondi 
solution \cite{TB}:

\bea
   A(r,t) &=& \frac{B'(r,t)}{\sqrt{1+C_1(r)}},
\label{eq:TB-A}
\eea
\bea
   B(r,t) &=& \left\{
     \begin{array}{ccl}
       \left(\frac{9C_2(r)}{4}\right)^{1/3}\left[t-C_3(r)\right]^{2/3} 
       & & \mbox{for}\,\,C_1(r)=0 \\
       \frac{C_2(r)}{2C_1(r)}\left(\cosh\eta-1\right),
       & \left( t-C_3(r)
        =\frac{C_2(r)}{2C_1^{3/2}(r)}\left(\sinh\eta-\eta\right)\right)
       & \mbox{for}\,\,C_1(r)>0 \\
       \frac{C_2(r)}{2|C_1(r)|}\left(1-\cos\eta\right), 
       & \left( t-C_3(r)
        =\frac{C_2(r)}{2|C_1^{3/2}(r)|}\left(\eta-\sin\eta\right)\right)
       & \mbox{for}\,\,C_1(r)<0 \\
       \end{array}
   \right. ,
\label{eq:TB-B}
\eea
\bea
   \kappa\rho(r,t) &=& \frac{C_2'(r)}{B^2B'},
\label{eq:TB-rho}
\eea
where $C_1(r)$, $C_2(r)$, and $C_3(r)$ are arbitrary functions of $r$ 
and a prime denotes the derivative with respect to $r$.
By taking $C_1(r)=c$, $C_2(r)=4r^{\alpha}/9$, and $C_3(r)=pr^{\alpha}$,
we obtain the Tolman-Bondi solution with self-similarity 
Eqs.(\ref{eq:FP-TB-c0})--(\ref{eq:FP-TB-c-}).

As for the calculation of linear perturbation, 
since we concentrate on the spherical modes of perturbation 
around a self-similar solutions 
it is enough to consider the linear perturbation of arbitrary functions, 
$C_1(r)$, $C_2(r)$, and $C_3(r)$.
The perturbed quantities, $\delta C_1(r)$, $\delta C_2(r)$, and 
$\delta C_3(r)$, can be expressed by a superposition of modes with 
different $\omega$ and taken in the following form,
\bea
 \delta C_1(r) &=& c_1r^\omega, \\
 \delta C_2(r) &=& c_2r^{\omega+\alpha}, \\
 \delta C_3(r) &=& c_3r^{\omega+\alpha}.
\eea

By coordinate transformation of $r$,
\bea
 r &\longrightarrow& r+F(r),
\eea
where $F(r)$ is an arbitrary function of $r$, 
we obtain the gauge mode of linear perturbation.
This function $F(r)$ also can be expressed by the superposition 
of modes with different $\omega$ in the form,
\bea
 F(r) &=& fr^{\omega+1}.
\label{eq:ctran}
\eea

\end{appendix}

%%%%%%%%%%%%%%%%%%%%%%%%%%%%%%%%%%%%%%%
%             References              %
%%%%%%%%%%%%%%%%%%%%%%%%%%%%%%%%%%%%%%%

\newpage
%%%%%%%%%%%%%%%%%%%%%%%%%%%%%%%%%%%%%%%
%               Figure                %
%%%%%%%%%%%%%%%%%%%%%%%%%%%%%%%%%%%%%%%

\begin{thebibliography}{99}
\bibitem{CT}
The RG equations also can be seen as the Einstein equations 
in the new coordinate system ($\tau,\xi,\theta,\phi$), where 
$\tau=(\ln t)/\alpha$, $\xi=r/t^{1/\alpha}$, 
$\L{A}(\xi,\tau)=t^{(1-\alpha)/\alpha}A(r,t)$, and 
$\L{B}(\xi,\tau)=t^{-1}B(r,t)$.
One might be puzzled by observing that
in this new coordinate system, the lapse function can be zero 
at $\xi^2\L{A}=\alpha^2$.
However, there are no physical meaning of this point because the Einstein 
equations are found to be regular at this point.
\end{thebibliography}
\end{document}